\documentclass[a4paper,11pt]{article}
\usepackage{jinstpub} % for details on the use of the package, please see the JINST-author-manual
\usepackage{lineno}
\usepackage{subcaption}
\usepackage[nolist,nohyperlinks]{acronym}
\usepackage{siunitx}
\usepackage{todonotes}

\title{Adaptive carrier signal synthesis for frequency-division multiplexed cryogenic detector readout}

\author[1]{T. Muscheid\note{Corresponding author.},}
\author{B. Keppner,}
\author{D. Crovo,}
\author{R. Gartmann,}
\author{M. Fuchs,}
\author{L. Scheller,}
\author{O. Sander,}
\author{and L. E. Ardila-Perez}

\affiliation{Institute for Data Processing and Electronics, Karlsruhe Institute of Technology, Karlsruhe, Germany}

\emailAdd{timo.muscheid@kit.edu}

\abstract{Frequency-domain multiplexing is a fundamental technology for readout of cryogenic detector arrays on a single transmission line. Synthesizing a large number of static carrier tones for resonator stimulation eventually drives the cryogenic low-noise amplifier beyond its P1dB, introducing detector signal distortion. Scaling array sizes to $\mathcal{O}(10^6)$ detector pixels therefore demands minimizing the total power on the transmission line by continuous tracking of the resonances, which requires the readout electronics to be capable of updating the carrier tones dynamically. 
This work presents an scalable FPGA-based implementation for carrier signal synthesis based on a numerically controlled oscillator and a polyphase filter bank, enabling real-time frequency adaptivity as well as frequency resolution of 4\,mHz to precisely tune the operating point, while simultaneously reducing resource usage compared to the current memory-based implementation.
}

\begin{acronym}
\acro{nco}[NCO]{Numerically-Controlled Oscillator}
\acro{ftw}[FTW]{Frequency Tuning Word}
\acro{tdm}[TDM]{Time-Division Multiplexing}
\acro{fdm}[FDM]{Frequency-Division Multiplexing}
\acro{lut}[LUT]{Look-Up Table}
\acro{awg}[AWG]{Arbitrary Waveform Generator}
\acro{fft}[FFT]{Fast-Fourier-Transform}
\acro{ifft}[IFFT]{Inverse Fast-Fourier-Transform}
\acro{dft}[DFT]{Discrete Fourier-Transform}
\acro{idft}[IDFT]{Inverse Discrete Fourier-Transform}
\acro{sfdr}[SFDR]{Spurious-Free Dynamic Range}
\acro{dsp}[DSP]{Digital Signal Processor}
\acro{fifo}[FIFO]{First In First Out}
\acro{pfb}[PFB]{Polyphase Filter Bank}
\acro{dac}[DAC]{Digital-to-Analog Converter}
\acro{ppc}[PPC]{Polyphase Channelizer}
\acro{pps}[PPS]{Polyphase Synthesizer}
\acro{lo}[LO]{Local Oscillator}
\acro{psd}[PSD]{Power Spectral Density}
\acro{sdr}[SDR]{Software-defined Radio}
\end{acronym}

\notoc

\begin{document}
\maketitle
%\flushbottom

% \newpage is required when adding \notoc in order to fix title page
\newpage

\section{Introduction}

%Readout of large-scale cryogenic detector arrays requires multiplexing schemes that transmit information from multiple sensor pixels via a single transmission line from the cryogenic environment to room temperature. 
%Early applications typically employed \ac{tdm} schemes by sampling detectors one after the other, resulting in detector bandwidth limitations. 
%Technological advancements established \ac{fdm} as a suitable readout approach for large scale detector arrays. High-bandwidth data converters enable sampling of multiple detectors simultaneously by stimulating the resonators and analyzing the respective carrier tone modulations to recover the detector signals. A high-level readout concept based on a \ac{sdr} is shown in figure \ref{fig:sdr_concept}.

\begin{figure}[htbp]
    \centering
    \includegraphics[width=0.7\textwidth]{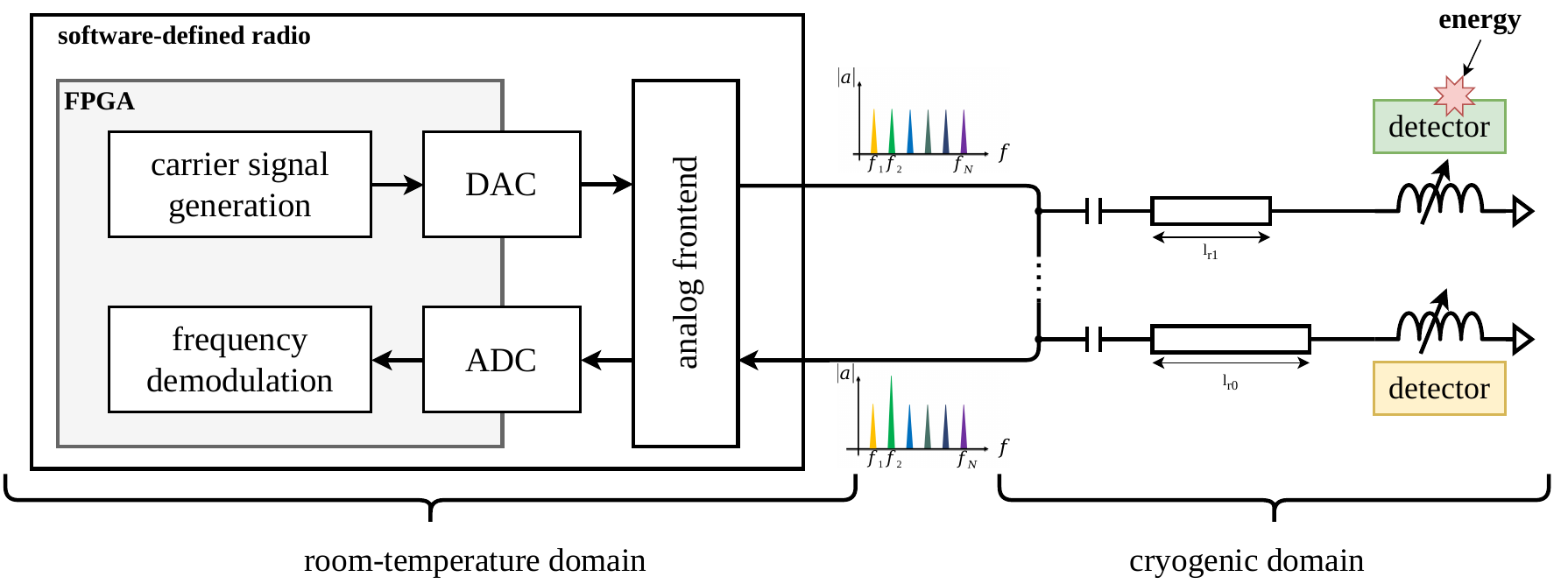}
    \caption{Readout of a \ac{fdm} detector array with a software-defined radio.}
    \label{fig:sdr_concept}
\end{figure}

%In \ac{fdm} systems, the sensor pixels act as a variable inductance as part of a resonator circuit. The detector signal is translated into a resonance frequency shift of the resonator this detector is coupled to. Operation of such a system requires the room-temperature instrumentation to generate a frequency comb containing the unmodulated resonance frequencies, here called \emph{carrier tones}, to stimulate the resonators. The detector signals and the resulting resonance frequency shifts are encoded into a modulation of the respective carrier tone. Frequency demodulation of this modulated frequency comb in the room-temperature readout electronics enable reconstruction of the raw detector signals.
Large-scale cryogenic detector arrays require dedicated readout electronics to benefit from the superior energy resolution of superconducting sensors. Using \ac{fdm} schemes, multiple detectors are operated on a single transmission line by inductively coupling the detector pixels to individual resonator circuits. The high-level readout concept is shown in figure \ref{fig:sdr_concept}, based on an \ac{sdr} for synthesizing a frequency comb with the resonance frequencies and analyzing the carrier signal modulation to reconstruct the detector signals.

In our current-generation systems \cite{Muscheid_2024, Muscheid_2026}, the carrier signals for resonator stimulation are synthesized with an \ac{awg}-like module. This module is implemented on the \ac{sdr}'s FPGA in form of a memory which stores the precomputed, reloadable time-domain samples of the comb. 
%While this \ac{awg} approach enables full control of the output signal shape, it shows two major disadvantages preventing its applicability for future, larger-scale use cases. 
However, the required memory scales linearly with the instantaneous bandwidth and the frequency resolution of the carrier tones, and dynamic frequency changes are not possible. 
These drawbacks limit further scalability of the readout electronics, both towards wider frequency bands as well as higher channel counts, which requires precise frequency tuning and real-time carrier tone adaptivity for advanced features like  tone tracking \cite{Yu_2023} or active resonator feedback \cite{Montgomery_2024}.
In this work, adaptive frequency comb generation based on a \ac{pps} is presented, eliminating the limitations of the \ac{awg} and offering a scalable, resource-efficient implementation. 
%In section \ref{sec:pps} the mathematical foundations behind this approach are explained, before the concrete module implementation is described in section \ref{sec:adaptive_frequency_comb}. The novel system is characterized and compared to the conventional approach in section \ref{sec:eval}. 

\section{Polyphase synthesis}
\label{sec:pps}
%Naively, adaptive frequency comb generation can be achieved by synthesizing many individual complex DC signals of the desired amplitude and phase, subsequently mixing them to their specific target frequency before summing all signal paths to create the wideband comb. Mathematically, this transformation is described by 

%\begin{equation}
%    \label{eq:parallel_approach}
%    y[n]= \sum_{k=0}^{N-1}x_k[n]*e^{j\omega_k[n]}.
%\end{equation}

%To fulfill the Nyquist theorem, the sampling rate of each input signal needs to be equal to the sampling rate of the combined wideband output. The input signal could also be generated in an interleaved form to reduce the number of high-frequency inputs, which requires additional upsampling modules before the mixing stage. However, this approach is not resource efficient, as the number of baseband input signals, mixers and addition operation scales linearly with the number of tones in the comb. 

The \ac{pps} processes an input data stream consisting of $M$ interleaved low-bandwidth baseband signals with an $M$-point \ac{idft} followed by a \ac{pfb} \cite{xilinx:xapp1161}. The \ac{idft} assigns the individual signals to a corresponding channel of the wideband comb, while the \ac{pfb} converts the baseband sampling rate to the output rate and filters spectral images. At the \ac{pfb} output, the data stream contains the wideband frequency spectrum with all desired carrier tones. Mathematically, the concept of the \ac{pfb} can be described by equation \ref{eq:pfb}. It is processing the result of the $M$-point \ac{idft} $x_n$ by interleaved FIR filters of order $F$ and combining the filtered data into a single data stream via an output commutator.

% \begin{equation}
%   \label{eq:dft}
%   x_n = \frac{1}{M}\sum_{k=0}^{M-1}X_k*e^{i 2 \pi \frac{k}{M} n}.
% \end{equation}

\begin{equation}
    \label{eq:pfb}
    y[M*n+l]= \sum_{k=0}^{F-1}x_{n-k}[l] \cdot b_{M*k + l} \qquad l \in \{0,1, ... ,M-1\}
\end{equation}

%\begin{figure}[htbp]
%    \centering
%    \includegraphics[width=0.45\textwidth]{figures/poly_synth.pdf}
%    \qquad
%    \includegraphics[width=0.45\textwidth]{example-image-a}
%    \caption{Left: Frequency comb generation based on a polyphase filter bank. Right: Output spectrum of an exemplary frequency comb.}
%    \label{fig:polyphase_synthesizer}
%\end{figure}

%While frequency comb analysis based on polyphase channelization is the most common approach used by readout electronics for \ac{fdm} cryogenic detector arrays, synthesis based on this method is at present only used rarely. Mathematically, synthesis and channelization requires the same operations in inverse order. In contrast to the \ac{ppc}, the intermediate signal at the output of the \ac{ifft} can directly be processed by the filter bank, without the need for additional reordering.

The number of channels in the frequency comb is adaptable to specific experimental requirements by appropriately parameterizing the value of $M$.
This parameter defines the number of \ac{pfb} channels and limits the maximum channel bandwidth $B_\mathrm{ch} \le \frac{f_s}{M}$, which can be further reduced by defining a custom cut-off frequency in the polyphase filter coefficients. Similarly, the order of the filter coefficients $F$ can be set to achieve the desired stop-band rejection. Both higher channel counts and better stop-band rejection result in increased resource consumption; hence, the \ac{pps} should be designed to match the requirements of the specific application.
With the \emph{xfft} and the \emph{fir\_compiler} IP-cores, AMD offers suitable implementations for this application, which can be customized to the specific application by defining the sampling frequencies, channel count, and filter coefficients appropriately. Since the \ac{idft} is implemented as an \ac{ifft} for resource efficiency reasons, the number of channels in the \ac{pps} is restricted to $M = 2^m$ with $m \in \mathbb{N}$.

\section{Adaptive frequency comb conversion}
\label{sec:adaptive_frequency_comb}
\begin{figure}[htbp]
    \centering
    \includegraphics[width=0.8\textwidth]{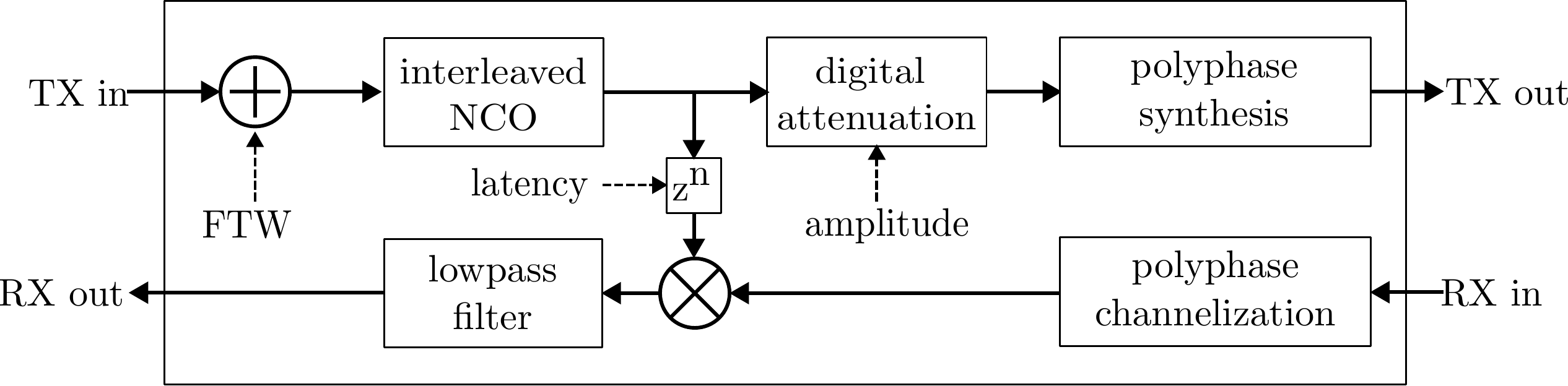}
    \caption{Architecture of the comb conversion module. Solid lines draw the signal flow through the module, while dashed lines indicate parameter values defined by the user at run-time.}
    \label{fig:comb_generator}
\end{figure}

The combination of the \ac{pps} with its inverse implementation for comb analysis, the \ac{ppc}, serves as the foundation for a fully self-contained comb conversion module. This comb conversion module is designed as a four-port device, creating a wideband comb (\emph{TX-out}) from user-defined frequencies and reconstructing the detector data (\emph{RX-out}) from the modulated comb (\emph{RX-in}). The tones in the frequency comb can be individually shifted with the signal at the \emph{TX-in} port. The block diagram of the module and its subsystems is shown in figure \ref{fig:comb_generator}. 

%A shared multi-channel \ac{nco} for up- and down-mixing is used. This not only reduces resource utilization but also eliminates the need for frequency calibration during down-mixing of the static carrier signals in the conventional system. In summary, the module offers efficient encapsulation of the bidirectional transfer between \ac{tdm} and baseband domain. 

As stated in section \ref{sec:pps}, the \ac{pps} requires interleaved baseband signals as input data. In this implementation, a \ac{lut}-based interleaved \ac{nco} is instantiated to synthesize a single tone for each channel, corresponding to the offset frequency of the respective bin center frequency of the \ac{pps}. These frequencies are configured using \SI{32}{\bit} \acp{ftw}, defining the phase increment of the \ac{nco}. While the initial \ac{ftw} is statically set by the user at run time, the optional \emph{TX-in} port enables adaptive tone generation by updating the \ac{nco}'s phase increment in real time.
%This \ac{nco} is designed around a \ac{lut}-based generation of a single complex frequency consisting of a \emph{sin} and \emph{cos} component. The \ac{nco}'s output frequency is set by a \ac{ftw}, defining the phase increment between two samples. By accumulating \acp{ftw} separately for multiple channels and storing the result in a ring buffer, individual frequencies for each channel can be synthesized in a \ac{tdm}-mode to match the expected input signal of the \ac{pps}. 
The \ac{nco} generates complex data samples for $M$ channels in \ac{tdm} with the output frequencies defined in the range $f_\mathrm{ch} \le \pm \frac{1}{2} \frac{f_s}{M}$. 
With the digital attenuation subsystem, each channel's full-scale \ac{nco} amplitude can be reduced individually to adapt the readout power to the resonator characteristics \cite{Wegner_2022}.
The resolution for channel frequency $\Delta f_{ch}$ and amplitude $\Delta A_{ch}$ are defined by:

\begin{equation}
    \label{eq:resolution}
    \Delta f_{ch} = \frac{f_{s}}{M*2^{32}}, \qquad \Delta A_{ch} = \frac{1}{2^{16}}.
\end{equation}

%The dynamic tone generation offers individual tuning of frequency and attenuation for each channel. For this configuration, the \acs{AXI} memory module presented in \cref{ch:firm:axi_memory} is used. The calibration options are described separately in the following paragraphs.
Since the comb conversion module consists of the \ac{pps} and \ac{ppc} with common parameters and filter coefficients, the downconverted interleaved frequencies at the \ac{ppc} output correspond exactly to the frequencies at the \ac{pps} input. This characteristic can be exploited by reusing the \ac{nco} signals as \acp{lo} for the second-stage downmixing. Here, the full scale amplitude of the \ac{nco} signal is used to preserve the dynamic range of the input stream, independent of the signal attenuation of the generated signal in TX-direction.
Due to signal propagation through the data converters, the analog frontend, and the cryogenic environment, the signal at the \emph{RX-in} port is delayed with respect to the \emph{TX-out} signal. To ensure down-conversion of the received signal with the correct \ac{lo} frequency, a \ac{fifo} buffer of variable length $N_d$ is implemented to delay the NCO signal accordingly. 
%This delay of the buffer can be configured at runtime to precisely match $N_d = \tau _d * f_s$, where $\tau _d$ defines the signal propagation time, which is measured by monitoring the two mixer input signals during a programmed frequency jump and counting the number of samples ($N_d$) between the TX frequency change and the corresponding RX response. 
To suppress aliasing and high-frequency noise, the downmixed signals are processed by a FIR filter with coefficients carefully computed at build-time to align the cut-off frequency with the expected detector bandwidth. After this processing step, the detector signals are successfully reconstructed and can be further processed by subsequent modules connected to the \emph{RX-out} port.
\newline
Runtime configuration of the \acp{ftw} and tone amplitudes as well as the latency $N_d$ is enabled by a network interface, allowing users to conveniently control the \ac{sdr} remotely via the custom \emph{cirque} Python API \cite{cirque2025}. The coefficients for the \ac{pfb} in the \ac{pps} and the \ac{ppc}, as well as for the low-pass filter after downmixing, are created using a MATLAB script at firmware build time. Depending on the desired number of channels, the detector bandwidth, and stopband rejection, this script automatically generates the coefficient files, which are loaded into the respective filter modules when creating the FPGA bitstream. This approach enables custom firmware images that exactly target the applications' requirements to maintain low resource consumption. 
%with four signal ports. The two baseband ports (RX-in and TX-out) are connected to the analog frontend. The baseband ports contain the received and the generated carrier combs. The two \ac{tdm} ports (TX-in and RX-out) are exposed to the firmware modules responsible for synthesizing and analyzing the signals. The \ac{tdm} RX-out port contains the down-mixed channelized complex $s_{21}$ data while the TX-in port is used for real-time shifting of the generated carrier signals. The TX-in port expects \acp{ftw}, which are added onto the user-configured channel \ac{ftw}. 

\section{Performance evaluation
%Evaluation
}
\label{sec:eval}

To evaluate the new approach for frequency comb synthesis, the ECHo-100k room-temperature readout electronics \cite{Muscheid_2024} has been utilized. The comb conversion module was implemented into the programmable logic of the ZynqUS+ MPSoC on the DTS-100G board \cite{Muscheid_2023}. In this test setup, one instance of the module is covering a frequency bandwidth of 500\,MHz with 32 channels, mixed up by the \ac{sdr}'s analog frontend to 4.4\,GHz.
Compared to the \ac{awg} approach, the comb conversion module improves the frequency resolution $\Delta f$ by several orders of magnitude. As defined by equation \ref{eq:resolution}, the frequency resolution of the comb conversion is dependent on the number of channels and the bit width of the \ac{nco} phase input, resulting in $\Delta f \approx 4\,mHz$ for this application. In contrast, the frequency resolution of the conventional \ac{awg} approach depends on the output sampling rate $f_s$ and the number of samples $N$ stored in the memory and is given by $\Delta f = \frac{f_s}{N}$, typically achieving frequency resolutions in the order of 1\,kHz . 
At the same time, the novel approach drastically reduces the memory utilization, enabling implementation of additional memory-intensive features at later stages in the signal processing chain. To achieve a frequency resolution of 1\,kHz in this test setup, the \ac{awg} module requires 16\,Mb memory. Instead, the \ac{pfb} and the \ac{ifft} utilizes \ac{dsp} slices, which are usually not the limiting type of resource in our systems. The required resources for the implemented configuration are listed in table \ref{tab:resources}.

\begin{table}[htbp]
\centering
\caption{Resource utilization for the comb conversion module with 500\,MHz of bandwidth distributed into 32 channels with 90\,dB stopband rejection.}
\label{tab:resources}
\smallskip
\begin{tabular}{l|c|c|c|c}
\hline
 &Memory&DSP&LUT&FF\\
\hline
NCO & 270 Kb & 2 & 257 & 564\\
PPS & 18 Kb & 38 & 1854 & 4345\\
PPC & 0 & 40 & 1754 & 4107\\
Mixer + Filter & 0 & 17 & 452 & 1192\\
\hline
XCZU19EG & 70.6 Mb & 1968 & 523k & 1045k\\
\hline
\end{tabular}
\end{table}

\begin{figure}[htbp]
    \centering
    \includegraphics[width=\textwidth]{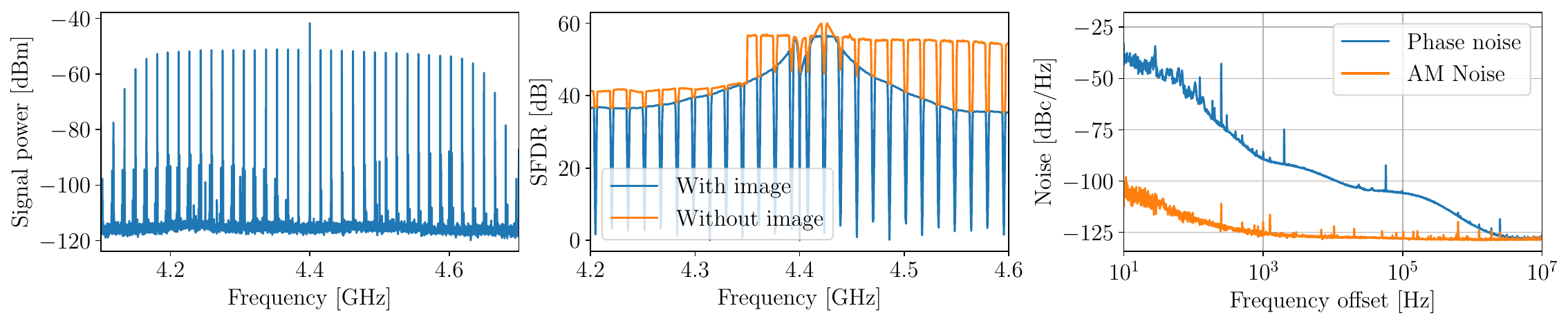}
    \caption{(a): Exemplary frequency comb with 32 tones generated by the \ac{pps}. (b): \ac{sfdr} showing the \ac{pfb} blind intervals and the image frequency as most prominent spur. (c): Noise \ac{psd} of a single tone at 4.52\,GHz.}
    \label{fig:spectral_shape}
\end{figure}

Evaluation of the synthesiszed signal is carried out with a spectral analyzer\footnote{Rohde\&Schwarz FSWP-50 with low phase-noise option FSWP-B61} connected to the transmission output port of the \ac{sdr}'s analog front-end. A frequency comb with 32 tones synthesized by the comb conversion module is displayed in figure \ref{fig:spectral_shape}a, successfully showing the applicability of the \ac{pps}. The spectrum clearly shows the desired tones centered around the \ac{lo} at 4.4\,GHz. The spurious signals in the lower sideband are crosstalk artifacts from the intermediate frequency, originating from the two-stage mixing scheme in the analog front-end \cite{Gartmann_2022} and also present in the spectrum of the \ac{awg} comb. To characterize the noise performance of the comb conversion module, a single tone was swept across the 500\,MHz spectrum and analyzed with the spectral analyzer to calculate the \ac{sfdr}. In the channel passbands, in \ac{sfdr} is consistently above 35\,dB, while the blind intervals between two neighboring channels and the frequency range close to the \ac{lo} are not suitable for carrier signal generation (see figure \ref{fig:spectral_shape}b).
Without the image frequency generated by the mixer due to IQ imperfections, the \ac{sfdr} is improved to 40\,dB, again revealing the crosstalk artifacts from the intermediate frequency. Additionally, the noise power spectral density was measured for a single tone at 4.52\,GHz, resulting in amplitude and phase noise below 90\,dBc/Hz for a frequency offset above 1\,kHz (see figure \ref{fig:spectral_shape}c).
%Regarding noise performance, both approaches are equally good and signal quality is limited by the converter noise.

\section{Design trade-offs and mitigation strategies
%Analysis and design trade-offs
%Discussion
}

The presented module for adaptive frequency comb synthesis is in many ways advantageous over the \ac{awg} implementation. It was implemented with the intention to update the actual resonance frequencies in real-time. For this use-case, combining the \ac{pps} and the \ac{pps} with common configuration to work with the same intermediate signals both in TX and RX directions is beneficial, as it allows fully automatic, low-latency system calibration. Given the excellent frequency resolution and the simple adaptation of the carrier tones by tuning the \acp{ftw}, the implemented comb conversion module fulfills all requirements to enable further scalability. 

However, because of algorithmic constraints and architectural decisions, the comb conversion module shows some restrictions regarding the output signal shape. Due to the nature of the \ac{nco}, which generates a pure complex frequency, the comb conversion module supports only a single tone per channel. Since resonator spacing of ten times the detector bandwidth is desired to suppress crosstalk effects \cite{Mates_2019}, the comb conversion module is typically able to match the detector design by appropriately parameterizing the sampling rate and the channel number of the \ac{pfb}. In edge cases with densely spaced resonators, synthesis of multiple tones per bin could be achieved by operating several \acp{nco} in parallel and superposing the signals before processing them with the \ac{pps}. A more critical limitation for detector designs with non-equidistantly spaced resonators is that no signals can be generated in the blind intervals between \ac{pfb} channels. These blind intervals can be covered by extending the module to an \emph{oversampled \ac{pfb}} with overlapping channels \cite{Smith_2021, Basha_2025} or by adding a second frequency-shifted \acp{pfb} in parallel.
Finally, to suppress image frequencies for \ac{sfdr} improvement (compare figure \ref{fig:spectral_shape}b), the I and Q components of the complex signal can be predistorted in phase and amplitude for trace length compensation and correction of component variations \cite{ad:an1039}. With the \ac{awg} and its full control of the signal shape, individually parameterizing I and Q is straightforward. A similar compensation concept can be implemented into the comb conversion module by extending the digital attenuation module to attenuate the baseband I and Q components with individual factors.

%With the \ac{awg}, it is possible to generate a single or a few tones with relatively high amplitude. The total power is limited by the bit width of the \ac{dac}, typically 16 bit, but can be freely distributed across the frequencies. 
%On the other hand, the amplitude of individual carriers is limited to $P/M$ in the \ac{pps}, independent of the actual number of generated tones but only on the number of channels $M$ of the \ac{pfb}.
%Under certain circumstances, i.e. multiplexers with densely or non-equidistantly spaced resonators, the \ac{awg} might be better suited and will therefore still be supported. 
%One could argue that the \ac{pps} is not the optimal choice for generating unmodulated frequency combs, as the underlying algorithm was developed to synthesize and analyze broadband carrier tones modulated by some kind of information, as is the case for the RX part of our system. In the present use case with one single sinusoidal unmodulated carrier tone per channel, the available channel bandwidth is not fully exploited.  

\section{Conclusion}
In this work, a self-contained comb conversion module for adaptive carrier signal synthesis has been implemented on a FPGA and validated under realistic test conditions. By combining polyphase synthesis with the established polyphase channelization algorithm, the module performs frequency demodulation automatically and enables real time carrier tone adaptation with sub-Hz frequency resolution while significantly reducing memory consumption compared to the conventional memory-based implementation. Its parametric design allows flexible configuration targeting specific applications, yielding resource-optimized implementations. Integrating the comb conversion module into next-generation readout electronics simplifies system calibration and paves the way for operation of even larger-scale detector arrays.

\newpage

%The \ac{pps} for adaptive carrier signal synthesis has been successfully implemented and evaluated under realistic test setup conditions. 
%In combination with the well-established \ac{ppc} module, a self-contained comb conversion module with automatic calibration for frequency demodulation was designed, simplifying the measurement setup procedure and the usability of the system. This module is able to synthesize parametric frequency combs with resolution better than 1\,Hz while simultaneously drastically reducing memory consumption. 
%With the \emph{TX-in} port, the module offers a modular way to adapt the carrier frequencies in real time by updating the \ac{ftw}, paving the way for advanced readout concepts.
%Compared to the conventional \ac{awg} approach, the system presented here exhibits limitations regarding signal shape, which can be solved by replacing or extending individual subsystems if needed.

%\appendix
%\section{Some title}
%Please always give a title also for appendices.

%\acknowledgments
%We acknowledge the support of the Doctoral School \emph{“Karlsruhe School of Elementary and Astroparticle Physics: Science and Technology”}

%This is the most common positions for acknowledgments. A macro is
%available to maintain the same layout and spelling of the heading.

%\paragraph{Note added.} This is also a good position for notes added
%after the paper has been written.

% Bibliography

%% [A] Recommended: using JHEP.bst file
\bibliographystyle{JHEP}
\bibliography{biblio.bib}

\end{document}